\documentclass[12pt]{report}

\begin{document}

\begin{center}
Abstract Submitted \\
for the APR05 Meeting of \\
The American Physical Society
\end{center}

\begin{flushright} 
Sorting Category: E10. (T)
\end{flushright}

\begin{quotation}
\begin{center}
{\bf The number of $J=0$ pairs in $^{44,46,48}$Ti}

LARRY ZAMICK, ALBERTO ESCUDEROS and ARAM MEKJIAN \\ (Rutgers University)
\end{center}

In the single $j$-shell, the configuration of an even--even Ti isotope consists
of 2 protons and $n$ neutrons. The $I=0$ wave function can be written as
$$
\psi=\sum_{J v} D(J,J v) [(j^2)^J_\pi (j^n)^J_\nu]^{I=0},
$$
where $v$ is the seniority quantum number. There are several states with 
isospin $T_{\rm min}=|(N-Z)/2|$, but only one with $T_{\rm max}=T_{\rm min}+2$.
By demanding that the $T_{\rm max}$ wave function be orthogonal to the $T_{\rm 
min}$ ones, we obtain the following relation involving a one-particle 
coefficient of fractional parentage:
$$
D(00)=\frac{n}{2j+1} \sum_J D(J,J v) (j^{n-1}(j v=1) j|j^n J) \sqrt{2J+1}.
$$
This leads to the following simple expressions for the number of $J=0$ $np$ 
pairs in these Ti isotopes:
\begin{itemize}
\item For $T=T_{\rm min}$, \hspace{5mm} \# of pairs $(J_{12}=0)=2|D(00)|^2/n$

\item For $T=T_{\rm max}$, \hspace{4mm} \# of pairs $(J_{12}=0)=2n|D(00)|^2=$

\hspace{6.75cm} $=$ \large$\frac{2n(2j+1-n)}{(2j+1)(n+1)}$
\end{itemize}
For $^{44}$Ti we have also the results for even $J_{12}$
$$
{\rm \#\ of}\ nn\ {\rm pairs}={\rm \#\ of}\ pp\ {\rm pairs}={\rm \#\ of}\ np\ 
{\rm pairs}=|D(J_{12},J_{12})|^2.
$$
\end{quotation}

\end{document}